\documentclass[conference]{IEEEtran}
\ifCLASSINFOpdf
\else
\fi

\usepackage{amsmath,amssymb}
\usepackage{mathrsfs}
\usepackage{graphicx}
\usepackage{amsthm}
\theoremstyle{definition}

\newtheorem*{theorem*}{Theorem}

\usepackage{txfonts}
\usepackage{comment}

\begin{document}
%
\title{Method to Design UF-OFDM Filter and its Analysis}

\author{\IEEEauthorblockN{Hirofumi Tsuda}
\IEEEauthorblockA{Department of Applied Mathematics and Physics\\
Graduate School of Informatics\\ Kyoto University\\
Kyoto, Japan\\
Email: tsuda.hirofumi.38u@st.kyoto-u.ac.jp}
\and
\IEEEauthorblockN{Ken Umeno}
\IEEEauthorblockA{Department of Applied Mathematics and Physics\\
Graduate School of Informatics\\ Kyoto University\\
Kyoto, Japan\\
Email: umeno.ken.8z@kyoto-u.ac.jp}}


%


\maketitle

\begin{abstract}
Orthogonal Frequency Division Multiplexing (OFDM) systems have been widely used as a communication system. In OFDM systems, there are two main problems. One of them is that OFDM signals have high Peak-to-Average Power Ratio (PAPR). The other problem is that OFDM signals have large side-lobes. In particular, to reduce side-lobes, Universal-Filtered OFDM (UF-OFDM) systems have been proposed. In this  paper, we show criteria for designing filters for UF-OFDM systems and a method to obtain the filter as a solution of an optimization problem. Further, we evaluate PAPR with UF-OFDM systems. Our filters have smaller side-lobes and lower Bit Error Rate than the Dolph-Chebyshev filter. However, the PAPR for signals with our filters and the Dolph-Chebyshev filter are higher PAPR than those with conventional OFDM signals.
\end{abstract}


%
\IEEEpeerreviewmaketitle

\section{Introduction}
Orthogonal Frequency Division Multiplexing (OFDM) systems have been widely used for broadband multicarrier communication since OFDM systems have a number of advantages. One of advantages is that OFDM system can deal with multipath fading. The reason for this is that OFDM signals are generated with inverse Fourier transformation. Another advantage is that OFDM systems can be adopted to multiple-input multiple-output (MIMO) channels. Due to these advantages, OFDM systems are used as some telecommunication standards. 

However, it is known that there are two main problems in OFDM systems. One of them is that OFDM systems have high Peak-to-Average Power Ratio (PAPR), which is the ratio between the maximum value of signal powers and the average value of them. For low values of the input powers, the output power grows approximately linear. However, for input signals with large power, the growth of the output power is not linear. Then, in-band distortion and out-of-band distortion are caused \cite{ofdm}. The effects of non-linear amplifiers are investigated in \cite{costa}. Therefore, techniques to reduce PAPR are demanded. There are many methods to reduce PAPR to avoid such problems \cite{armstrong}-\cite{airy}. Further, the distribution of the PAPR and performances of methods are investigated in \cite{ochiai} \cite{clip}.

The other main problem is that OFDM signals have large side-lobes \cite{filterbank}. This problem leads to leakage of signal powers among the bands of different users. Therefore, there are some improved and proposed systems to solve this problem. One of the systems is a Universal-Filtered OFDM (UF-OFDM) system \cite{ufofdm}. In UF-OFDM systems, band-bass filters are used to reduce side-lobes. Therefore, methods to design efficient filters are demanded. There are some investigations about improving UF-OFDM systems \cite{cancel} \cite{emission}.

In this paper, we show a method to design filters for UF-OFDM systems. As a signal processing technique for recovering symbols, we consider Zero Forcing equalization. Then, we derive the sufficient condition for increase of the Signal-to-Noise Ratio (SNR). From this condition, we obtain the optimization problem to increase the SNR and reduce side-lobes. Therefore, the filter for UF-OFDM systems is obtained as a solution of the optimization problem. Further, we evaluate the PAPR with UF-OFDM systems and show the relation between PAPR and SNR. Then, we show the numerical results about power spectrum density, Bit Error Rate (BER) and PAPR with filters obtained from the optimization problem. 
 
\section*{Mathematical Notation}
In what follows, we use the Fourier transformation as a discrete Fourier transformation
\begin{equation}
\mathcal{F}_{\mathbf{x}}(\omega) = \sum_{n}x_n\exp(-jn\omega),
\end{equation}
where $\mathbf{x} \in \mathbb{C}^p$, $x_n$ is the $n$-th element of $\mathbf{x}$, $j$ is the unit imaginary number and $p$ is a positive integer. For convenience, we write the discrete Fourier transformation of $\mathbf{x}$ as its capital, that is,
\begin{equation}
X(\omega) = \mathcal{F}_{\mathbf{x}}(\omega).
\end{equation}

Further, we write the complex Gaussian distribution whose average and variance are $\mu$ and $\sigma^2$ as
$\mathcal{N}(\mu,\sigma^2)$. Note that their real part and imaginary part obey the Gaussian distribution whose variance is $\sigma^2/2$, respectively. 

\section{UF-OFDM Model}
In this section, we fix our model used thorough this paper and mathematical symbols that will be used in the following sections. We consider Zero Padding (ZP) UF-OFDM systems. For more details of this model, we refer the reader to \cite{bell}. Let us define $A_k$ as a symbol transmitted by the $k$-th carrier. We assume that each $A_k$ is independent and $\operatorname{E}\{A_iA_k^*\} = \delta_{ik}$, where $\operatorname{E}\{X\}$ is the average of $X$, $z^*$ is the complex conjugate of $z$ and
\begin{equation}
\delta_{ik} = \left\{ \begin{array}{cc}
1 & i=k\\
0 & i \neq k
\end{array} \right. .
\end{equation}
Then, a discrete OFDM signal $x_n$ is written as

\begin{equation}
x_n = \sum_{k \in J}\frac{A_k}{\sqrt{M}}\exp\left(2 \pi j \frac{kn}{M}\right)\Pi\left(\frac{n}{M}-\frac{1}{2}\right) \hspace{3mm}(n=0,1,\ldots,M-1), 
\label{eq:ifft}
\end{equation}
where $J \subseteq \{0,1,\ldots,M-1\}$ is the set of the numbers of used carriers and $\Pi(x)$ is defined as \cite{theory}
\begin{equation}
\Pi(x) = \left\{ \begin{array}{cc}
1 & -\frac{1}{2} \leq x < \frac{1}{2}\\
0 & \mbox{otherwise}
\end{array}
\right. .
\end{equation}
Note that the duration of symbols is $M$. In particular, we denote $K$ by the number of elements in the set $J$. Equation (\ref{eq:ifft}) is implemented with the Inverse Fast Fourier Transformation (IFFT). In UF-OFDM systems, discrete OFDM signals are filtered to reduce their side-lobes. 
A filter $\mathbf{f} \in \mathbb{C}^N$ is written as
\begin{equation}
\mathbf{f} = \left[\begin{array}{cccc}
f_0 & f_1 & \cdots & f_{N-1}
\end{array}
 \right]^\top,
\end{equation} 
where $\mathbf{x}^\top$ is the transpose of $\mathbf{x}$. We assume that $f_0 \neq 0$ and $f_{N-1} \neq 0$. This assumption is equivalent to that the convolved signal has the length $M+N-1$. 
When the signal $x_n$ is filtered by $\mathbf{f}$, the output signal $y_n$ is written as 
\begin{equation}
y_n = \sum_{m=0}^{N-1}f_m\sum_{k \in J}\frac{A_k}{\sqrt{M}}\exp\left(2 \pi j \frac{k(n-m)}{M}\right)\Pi\left(\frac{n-m}{M}-\frac{1}{2}\right)
\label{eq:conv_t}
\end{equation}
for $n=0,1,\ldots,M+N-2$. Note that the duration of symbols becomes $M$ to $M+N-1$ due to the filtering. Equation (\ref{eq:conv_t}) is often written as a linear system with a Toeplitz matrix and a Discrete Fourier Transformation (DFT) matrix. 

In our model, we apply a zero padding technique to avoid intersymbol interference. Let us define $D$ as the length of zero paddings. Then, the transmitted signal $\tilde{y}_n$ is written as
\begin{equation}
\tilde{y}_n = \left\{ \begin{array}{cc}
y_n & 0 \leq n < M+N-1\\
0 & M+N-1 \leq n < M+N+D-1
\end{array} \right.
\label{eq:zeropadded}
\end{equation}

In wireless communication systems, fading effects should be considered. In OFDM systems, fading effects are expressed as a discrete convolution. Let us define $\mathbf{h} \in \mathbb{C}^L$ as
\begin{equation}
\mathbf{h} = \left[\begin{array}{cccc}
h_0 & h_1 & \cdots & h_{L-1}
\end{array}
 \right]^\top,
\end{equation}
where $L$ is the length of the fading $\mathbf{h}$.
Further, we assume that $L-1 \leq D$. This assumption is equivalent to that there we have the knowledge about the length of the fading effects. This assumption is often used \cite{zp}. 
Then, the received signal $r_n$ is written as
\begin{equation}
r_n = \sum_{l=0}^{L-1}h_l\tilde{y}_{n-l}+v_n\hspace{2mm}(n=0,1,\ldots,M+N+D-2),
\label{eq:r}
\end{equation}
where $v_n$ is additive white Gaussian noise (AWGN).

To estimate symbols, zero padding techniques are used. Then, the length of $r_n$ is extends to $2M$. Therefore, the zero padded signal $\tilde{r}_n$ is written as
\begin{equation}
\tilde{r}_n = \left\{ \begin{array}{cc}
r_n & 0 \leq n < M+N+D-1\\
0 & M+N+D-1 \leq n < 2M
\end{array} \right. .
\label{eq:r_zero}
\end{equation}

From Eqs. (\ref{eq:r}) and (\ref{eq:r_zero}), their Fourier transformations are written as
\begin{equation}
R(\omega) = \tilde{R}(\omega) = F(\omega)H(\omega)X(\omega) + V(\omega).
\label{eq:t_r}
\end{equation}
In particular, $X\left(2 \pi \frac{k}{M}\right)\hspace{2mm}(k \in J)$ is written as
\begin{equation}
X\left(2 \pi \frac{k}{M}\right) = \sqrt{M}A_k.
\end{equation}
Therefore, Eq. (\ref{eq:t_r}) is rewritten as
\begin{equation}
\frac{1}{\sqrt{M}}R\left(2 \pi \frac{k}{M}\right) = A_kF\left(2 \pi \frac{k}{M}\right)H\left(2 \pi \frac{k}{M}\right) + \frac{1}{\sqrt{M}}V\left(2 \pi \frac{k}{M}\right).
\label{eq:desire}
\end{equation}
The quantity $\tilde{R}\left(2\pi\frac{k}{M}\right)$ can be obtained with the $2M$-Fast Fourier Transformation and a down sampling technique \cite{bell}.
We can calculate and obtain $F(\omega)$ since we know $\mathbf{f}$. With some techniques, for example pilot symbols, $H(\omega)$ can be estimated and obtained. Therefore, from Eq. (\ref{eq:desire}), the symbol $A_k$ can be recovered with multiplication of the inverse of $H(\omega)$. Then, when the $H(\omega)$ is given and fixed, the SNR about the symbol $A_k$ is written as
\begin{equation}
\operatorname{SNR}_k = \sqrt{\frac{M}{M+N+D-1}}\frac{\left|F\left(2 \pi \frac{k}{M}\right)H\left(2 \pi \frac{k}{M}\right)\right|}{\sigma_n},
\label{eq:snr}
\end{equation}
where $\sigma_n$ satisfies $\operatorname{E}\{v^*_nv_n\} = \sigma^2_n$.
From Eq. (\ref{eq:snr}), SNR gets higher as $F\left(2 \pi \frac{k}{M}\right)$ gets larger. From this result, the large $F\left(2 \pi \frac{k}{M}\right)$ is demanded for high SNR.

Note that UF-OFDM systems are different from OFDM systems in a sense of SNR. In UF-OFDM systems, the duration of a symbol becomes longer than original signals with zero padding techniques and filtering. Then, the duration of a symbol with UF-OFDM system is longer than that with conventional OFDM systems. As seen in Eq. (\ref{eq:snr}), the SNR in UF-OFDM systems has the term $\sqrt{\frac{M}{M+N+D-1}}$, which is related to the length of filters $N$ and do not appear in conventional OFDM systems. From this result, UF-OFDM systems are different from OFDM systems in a sense of SNR.

\section{Optimization Problem for Designing Filter}
In this section, we show the criteria for designing filters. We focus on Bit Error Rate (BER) and side-lobes.
For convenience, let us define the $n$-th element of $\mathbf{g}$ as
\begin{equation}
g_n = \sum_{m}f_{n+m}f^*_m.
\label{eq:gf}
\end{equation}
Note that $\mathbf{g}$ is written as
\begin{equation}
\mathbf{g} = [\begin{array}{cccc}
g_{-N+1} & g_{-N+2} & \cdots & g_{N-1}
\end{array}]^\top
\end{equation}
It is known that the Fourier transformation of $\mathbf{g}$ is written as
\begin{equation}
\mathcal{F}_{\mathbf{g}}(\omega) = \sum_{n=-N+1}^{N-1}g_n\exp(-jn\omega) = |F(\omega)|^2.
\label{eq:power_f}
\end{equation}
From Eq. (\ref{eq:power_f}), $|F(\omega)|^2$ is expressed as the inner product of the two vectors, $\mathbf{g}$ and the vector whose $n$-th element is $\exp(-jn\omega)$. Further, when the vector $\mathbf{g}$ is given, we can obtain the original vector $\mathbf{f}$ with a spectral factorization method \cite{analysis}.

\subsection{Condition}
First, we consider the condition that filters have to satisfy. We assume that the average power of signals is conserved through the filter convolution. This assumption is written as
\begin{equation}
\frac{1}{M}\sum_{n=0}^{M-1}\operatorname{E}\{|x_n|^2\} = \frac{1}{M+N-1}\sum_{n=0}^{M+N-1}\operatorname{E}\{|y_n|^2\},
\label{eq:cond}
\end{equation}
where $\operatorname{E}\{X\}$ is the average of $X$. With $\mathbf{g}$, this condition is written as \cite{emission}
\begin{equation}
K(M+N-1) = \mathbf{b}_J^\top \mathbf{g},
\label{eq:cond_g}
\end{equation}
where $K$ is the number of the elements of $J$ and $\mathbf{b}_J$ is the vector written as
\begin{equation}
\begin{split}
\mathbf{b}_J & = \left[\begin{array}{cccc}
b_{J,-N+1} & b_{J,-N+2} & \cdots & b_{J,N-1}
\end{array}
\right],\\
b_{J,n} &= \sum_{k \in J}(M - |n|)\exp\left(-2 \pi j \frac{kn}{M}\right).
\end{split}
\label{eq:b_j}
\end{equation}
Note that Eq. (\ref{eq:cond_g}) is invariant under the following transformations:
\begin{itemize}
\item each carrier $k \rightarrow k + s$ and
\item the $n$-th element of the filter $\displaystyle f_n \rightarrow f_n\exp\left(2 \pi j \frac{sn}{M}\right)$ 
\end{itemize}
for any real value $s$. Therefore, when $J$ consists of $K$ consecutive integers\footnote{For example, $\{M-2,M-1,0,1,2,3\}$ is allowed.}, we can assume that the frequencies of carriers are symmetric around $\omega=0$, where $\omega$ is an angular frequency. This assumption is realized with the operation $k \rightarrow k + s$. Then, it is sufficient to design the filter whose power spectrum is symmetric around $\omega=0$. Therefore, we treat $\mathbf{f} \in \mathbb{R}^N$, and rewrite $\mathbf{g}$ as
\begin{equation}
\mathbf{g} = [\begin{array}{cccc}
g_{0} & g_{1} & \cdots & g_{N-1}
\end{array}]^\top
\end{equation}
since $\{g_n\}$ is symmetric around $n=0$. 
\subsection{Optimization Problem}
First, we divide the region $[0,\pi)$ into three regions as follows
\begin{equation}
[0,\pi) = \Omega_p \cup \Omega_t \cup \Omega_s,
\end{equation}
where $\Omega_p, \Omega_t$ and $\Omega_s$ are the regions of the passband, the transition-band and stop-band, respectively. We define the set of frequencies of carriers as $\Omega_{c}$, which is the frequency shifted set of $J$.

In Section II, we have shown that the large $F\left(\omega_c\right)\hspace{2mm}(\omega_c \in \Omega_c)$ is demanded for high SNR. As criteria of designing filters, high SNR and the rapid decay of side-lobes are demanded. Therefore, we obtain the following optimization problem
\begin{equation*}
\begin{split}
(P') \hspace{3mm}& \min \hspace{2mm} t_1 - \lambda t_2 \\
\mbox{s.t.}\hspace{2mm}& \left|F\left( \omega \right)\right|^2\operatorname{E}\{|X(\omega)|^2\} \leq t_1 \hspace{3mm}(\omega \in \Omega_s)\\
&\left|F\left(\omega_c\right)\right|^2 \geq t_2 \hspace{3mm}(\omega_c \in \Omega_c),\\
& K(M+N-1) = \mathbf{b}_c^\top \mathbf{g},\\
& \mathcal{F}_{\mathbf{g}}(\omega) \geq 0, \hspace{2mm}\omega \in [0, \pi],\\
& t_1,t_2 \geq 0,
\end{split}
\end{equation*}
where $\lambda$ is the positive weight parameter, $\operatorname{E}\{|X(\omega)|^2\}$ is written as
\begin{equation}
\begin{split}
\operatorname{E}\{|X(\omega)|^2\} &= \frac{1}{M}\sum_{\omega_c \in \Omega_c} \alpha\left(\omega_c - \omega\right)^2,\\
\alpha(\omega) &= \left\{ \begin{array}{c c}
M & \omega = 2n\pi\hspace{2mm}(n \in \mathbb{Z})\\
\sin\left(\frac{M\omega}{2}\right)/\sin\left(\frac{\omega}{2}\right) & \mbox{otherwise}
\end{array} \right.
\end{split}
\end{equation}
and $\mathbf{b}_c \in \mathbb{R}^{N}$ is the vector whose $n$-th element is defined as
\begin{equation}
b_{c,n} = \left\{\begin{array}{cc}
2\sum_{\omega_c \in \Omega_c}(M - |n|)\cos\left(n\omega_c\right) & n=1,2,\ldots,N-1\\
MN & n=0
\end{array} \right. .
\end{equation}
In the above equations, we have used the assumption that each $A_k$ is independent and $\operatorname{E}\{A_iA_k^*\}=\delta_{ik}$, which has been made in Section II.
In problem $(P')$, the variables are $\mathbf{g}$, $t_1$ and $t_2$. It is expected that the side-lobes get lower as the lambda becomes smaller. Besides, it is expected that the SNR gets larger as the lambda becomes larger. 

In the problem $(P')$, the first constraints are about side-lobes. The second constraints are about desired signals, which have been discussed in Section II. The third constraints are about the conservation of the signal power. The fourth constraints are necessary to satisfy Eq. (\ref{eq:gf}) \cite{sdp}. 
Note that the vector $\mathbf{b}_c$ is the same as $\mathbf{b}_J$ defined in Eq. (\ref{eq:b_j}) when the set $\Omega_c$ consists of the angular frequencies $2 \pi k/M$, where $k$ is the number carriers. In the problem $(P')$, it is not straightforward to handle the first and fourth constraints since $\Omega_s$ and $[0,\pi]$ are regions. To overcome these obstacles, we transform the problem.

One method to overcome the obstacles in the first constraints is to transform the constraints into discrete constraints \cite{sdp} \cite{linear}. With this method, the first constraints in the problem $(P')$ is replaced with
\begin{equation}
\left|F\left( \omega_i \right)\right|^2\operatorname{E}\{|X(\omega_i)|^2\} \leq t_1 \hspace{3mm}i=1,2,\ldots,S,
\end{equation}   
where $\omega_i\hspace{2mm}(i=1,2,\ldots,S)$ is the element in the region $\Omega_s$. In this paper, $S=15N$ is chosen. Note that how to choose the number $S$ is discussed in \cite{sdp} \cite{linear} \cite{fir}.

It is shown in \cite{sdp} that the fourth constraint is satisfied if there is a symmetric matrix $P$ which satisfies the following inequality
\begin{equation}
\left[\begin{array}{cc}
P-A^\top P A & C^\top - A^\top PB\\
C - B^\top PA & D+D^\top-B^\top PB
\end{array}
\right] \succcurlyeq 0,
\end{equation}
where 
\begin{equation}
\begin{split}
A &= \left[\begin{array}{ccccc}
0 & 0 & & \cdots & 0\\
1 & 0 &  & \cdots & 0\\
  & 1 &  &        &\\
\vdots &  & \ddots & \ddots & \vdots\\
0 & & & 1 & 0
\end{array} \right],\hspace{2mm} B=\left[\begin{array}{c}
1\\
0\\
\vdots\\
0
\end{array}\right],\\
C &= \left[\begin{array}{cccc}
g_1 & g_2 & \cdots & g_{N-1}
\end{array} \right], \hspace{2mm} D=\frac{1}{2}g_0
\end{split}
\end{equation}
and $X \succcurlyeq 0$ means that $X$ is a positive semidefinite matrix. 
With these methods, the problem $(P')$ is rewritten as
\begin{equation*}
\begin{split}
(P) \hspace{3mm}& \min \hspace{2mm} t_1 - \lambda t_2 \\
\mbox{s.t.}\hspace{2mm}& \left|F\left( \omega_i \right)\right|^2\operatorname{E}\{|X(\omega_i)|^2\} \leq t_1 \hspace{3mm}(\omega_i \in \Omega_s,i=1,2,\ldots,S)\\
&\left|F\left(\omega_c\right)\right|^2 \geq t_2 \hspace{3mm}(\omega_c \in \Omega_c),\\
& K(M+N-1) = \mathbf{b}_c^\top \mathbf{g},\\
& \left[\begin{array}{cc}
P-A^\top P A & C^\top - A^\top PB\\
C - B^\top PA & D+D^\top-B^\top PB
\end{array}
\right] \succcurlyeq 0,\\
& t_1,t_2 \geq 0, P \in \mathbb{S}^{N-1},
\end{split}
\end{equation*}
where $\mathbf{S}^n$ is the set of $n \times n$ symmetric matrices. 
Note that the variables in the problem $(P)$ are $\mathbf{g}, P, t_1$ and $t_2$. Further, the problem $(P)$ is convex.

\section{Peak-to-Average Power Ratio Analysis}
In Section III, we have shown the criteria for designing filters and the optimization problem. In this section, we discuss the PAPR of UF-OFDM signals and show the trade-off between PAPR and BER.

We make the following assumptions:
\begin{enumerate}
\item Continuous OFDM signals are regarded as a Gaussian process when the time $t$ is fixed.
\item Continuous UF-OFDM signals are regarded as a Gaussian process when the time $t$ is fixed.
\item Continuous OFDM signals are perfectly reconstructed from their discrete signals that are sampled.
\end{enumerate}
The first assumption is often used \cite{ochiai} \cite{clip}. This assumption is based on the central limit theorem. To analyse PAPR of UF-OFDM signals in a same way for OFDM signals, we make the second assumption.
About the last assumption, we assume that the following formula is satisfied:
\begin{equation}
\exp\left(2 \pi j \frac{k}{T}t\right) = \sum_{n=-\infty}^\infty \exp\left(2 \pi j \frac{kn}{M}\right)\Pi\left(\frac{n}{M}-\frac{1}{2}\right)\mbox{sinc}\left(\pi\left(\frac{t}{\Delta t} - n\right)\right)
\label{eq:sam_thm}
\end{equation}
for $0 \leq t \leq T$, where $\mbox{sinc}(x)$ is the sinc function, $T$ and $\Delta t$ are the symbol duration and the sampling time which satisfy $T = M\Delta t$. This formula is based on the sampling theorem. Note that Eq. (\ref{eq:sam_thm}) is not always satisfied since OFDM signals are not strictly band-limited due to the rectangular pulse $\Pi(t)$ (see Eq. (\ref{eq:ifft})) \cite{filterbank}. Therefore, Eq. (\ref{eq:sam_thm}) is an approximation.

The definition of the PAPR $\mathcal{P}$ is written as
\begin{equation}
\mathcal{P} = \frac{\max_{0 \leq t \leq T}|s(t)|^2}{P_{\operatorname{av}}},
\end{equation}
where $s(t)$ is a baseband signal and $P_{\operatorname{av}}$ is the average power of baseband signals. Note that PAPR is a random variable since signals are regarded as random variables. In this paper, it is called that the PAPR becomes high when the PAPR tends to have a large value.

From Eq. (\ref{eq:cond}) and the previously made assumption that the average power of signals are conserved through the filter convolution, the average power of UF-OFDM signals is written as
\begin{equation}
P_{\operatorname{av}} = \frac{K}{M}.
\end{equation}
Therefore, instead of $y_n$, we consider the following signal $\bar{y}_n$,
\begin{equation}
\begin{split}
\bar{y}_n &= \frac{y_n}{\sqrt{P_{\operatorname{av}}}}\\
 &=\sum_{m=0}^{N-1}f_m\sum_{k \in J}\frac{A_k}{\sqrt{K}}\exp\left(2 \pi j \frac{k(m-n)}{M}\right)\Pi\left(\frac{m-n}{M}-\frac{1}{2}\right).
\end{split}
\end{equation}
Let us consider the continuous UF-OFDM signal. We fix $t \in \left[T(N-1)/M,T\right]$. Note that the symbol duration of UF-OFDM signals is $T\times (M+N-1)/M$. From the sampling theorem, the continuous UF-OFDM signal is written as
\begin{equation}
\bar{y}(t) = \sum_{n=-\infty}^{\infty}\bar{y}_n\mbox{sinc}\left(\pi\left(\frac{t}{\Delta t} - n\right)\right).
\label{eq:uf_con}
\end{equation}
With Eq. (\ref{eq:sam_thm}), Eq. (\ref{eq:uf_con}) is rewritten as
\begin{equation}
\begin{split}
\bar{y}(t) &= \frac{1}{\sqrt{K}}\sum_{n=0}^{N-1}\sum_{k \in J}A_kf_n\exp\left(2 \pi j \frac{k}{T}\left(t-\frac{nT}{M}\right)\right)\\
&= \frac{1}{\sqrt{K}}\sum_{k \in J}A_kF\left(2\pi\frac{k}{M}\right)\exp\left(2 \pi j \frac{k}{T}t\right)
\end{split}
\end{equation}
for $t \in \left[T(N-1)/M,T\right]$. Note that $\bar{y}(t)$ is a random value since $A_k$ is a random value. The following formulas are satisfied:
\begin{equation}
\operatorname{E}\{\bar{y}(t)\}=0,\hspace{3mm}\operatorname{E}\{|\bar{y}(t)|^2\}=\frac{1}{K}\sum_{k \in J}\left|F\left(2\pi\frac{k}{M}\right)\right|^2.
\end{equation}
We denote $\tilde{\sigma}^2$ by $\sum_{k \in J}\left|F\left(2\pi\frac{k}{M}\right)\right|^2/K$.

In OFDM signals, the covariance of signals equals $1$. Therefore, from assumptions 2 and 3, signals of UF-OFDM $\bar{y}(t)$ and ones of OFDM $x(t)$ obey $\mathcal{N}\left(0,\tilde{\sigma}^2\right)$ and $\mathcal{N}(0,1)$, respectively. Therefore, their amplitudes obey the Rayleigh distributions whose scale parameters equal $\tilde{\sigma}/\sqrt{2}$ and $1/\sqrt{2}$, respectively. From this result, it is expected that the PAPR for UF-OFDM signals becomes higher as the parameter $\tilde{\sigma}$ gets larger. As seen in Section II, the parameter $\tilde{\sigma}$ relates to desired signals and large $\tilde{\sigma}$ is necessary for high SNR. Therefore, it is expected that the PAPR for UF-OFDM signals becomes higher as SNR gets higher.

\section{Numerical Results}
In this section, we show the BER, PAPR, and the power spectrum of filters obtained from the problem $(P)$. In our simulations, the parameters, the size of IFFT $M=128$, the length of filter $N=16$, the size of the zero-padding $D=16$ and the length of fading channels $L=12$ are chosen. Further, we assume that each element of the fading effect $h_n$ obeys $N(0,1/L)$. As carrier frequencies, we set $J=\{4,5,\ldots,19\}$. With frequency shifts, we design filters on the region $[0,\pi)$. We set the regions, $\Omega_p$, $\Omega_t$ and $\Omega_s$ as follows
\begin{equation}
\Omega_p = [0,17\pi/256), \Omega_t = [17\pi/256,17\pi/64), \Omega_s = [7\pi/64,\pi).
\end{equation}

With the above parameters, we obtain filters from the problem $(P)$ for various $\lambda$. To solve the problem $(P)$, we use the matlab package, CVX \cite{cvx}. The filter $\mathbf{f}$ is obtained from the solution of the problem $(P)$, $\mathbf{g}$ with spectral factorization methods \cite{fir}. We compare these filters with Dolph-Chebyshev filter whose side-lobe level is $-45$ dB. In each simulation, the modulation scheme is the QPSK modulation. In each figure, ``Our filters" means the filters obtained from the problem $(P)$ with various $\lambda$, ``D-C" means the Dolph-Chebyshev filter.

\begin{figure}[htbp]
\centering
\includegraphics[width=3.2in]{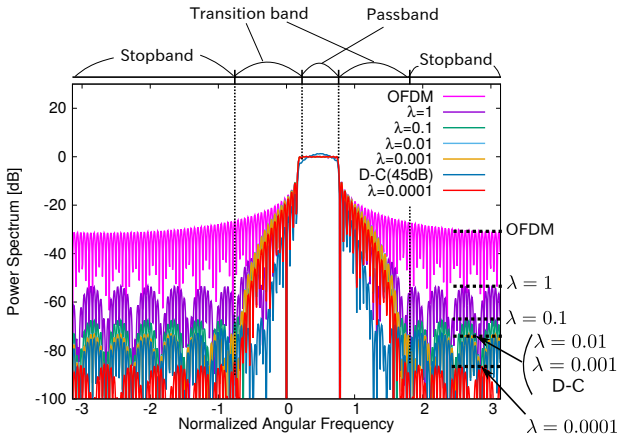}
\caption{Power Spectrum for UF-OFDM and OFDM systems}
\label{fig:spec}
\end{figure}

Figure \ref{fig:spec} shows the power spectrum of each filter. With filters obtained from the problem $(P)$, the side-lobe decreases as the weight parameter $\lambda$ gets smaller. In particular, with the filter whose parameter is $\lambda=0.0001$, its side-lobe is nearly $-90$ dB and lower than one of the Dolph-Chebyshev filter. It is known that side-lobe reductions in order of $-70$ dB are mandatory for broadcasting systems \cite{theory}. We observe that the filter whose parameter is $\lambda=0.0001$ is appropriate for broadcasting systems. Further, the side-lobe tends to be increased by the PAPR reduction schemes \cite{overview}. Thus, the filter whose parameter is $\lambda=0.0001$ may be still appropriate after processes of PAPR reductions.

\begin{figure}[htbp]
\centering
\includegraphics[width=3in]{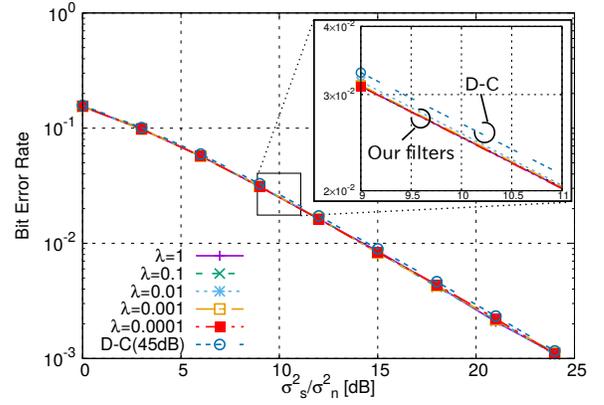}
\caption{Bit Error Rate of UF-OFDM with various filters}
\label{fig:ber}
\end{figure}

Figure \ref{fig:ber} shows the BER with various filters. We assume that the receiver has the perfect knowledge about the fading function $H(\omega)$. Then, the receiver can estimate the symbol $A_k$ with the knowledge of $H(\omega)$ (see Eq. (\ref{eq:desire})). In Fig. \ref{fig:ber},  $\sigma^2_s$ is the energy per bit before filtering and $\sigma^2_n$ is the variance of the AWGN channel, which has been defined in Eq. (\ref{eq:snr}). Note that the energy per bit of transmitted signals is $\frac{M+N-1}{M}\sigma^2_s$. The BER for our filters is lower than that for the Dolph-Chebyshev filter. In designing filters with the problem $(P)$, we increase the power of all desired signals. However, in the Dolph-Chebyshev filter, the power of all desired signals is not equal and there are some desired signals whose power is low. Therefore, it is conceivable that this difference at each power spectrum causes the difference of BER.
 
\begin{figure}[htbp]
\centering
\includegraphics[width=3in]{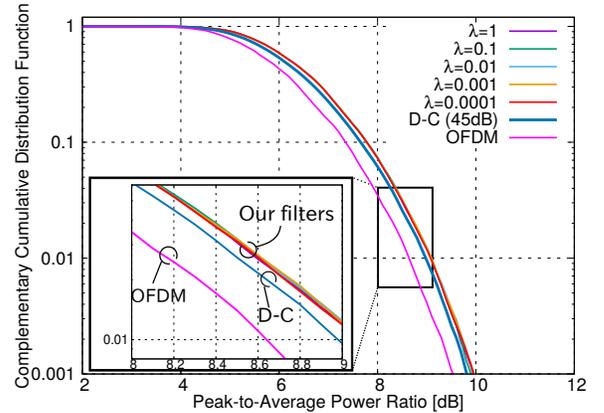}
\caption{Peak-to-Average Power Ratio for UF-OFDM and OFDM systems}
\label{fig:papr}
\end{figure}

Figure \ref{fig:papr} shows the PAPR with each filter. To compare equally, there is no zero paddings in UF-OFDM systems, that is, UF-OFDM signals are uniformly filled in the symbol duration. In calculating the PAPR, we set the $M=128$ points in each symbol duration and obtain each amplitude in the points. Then, the maximum amplitude is regarded as the maximum value in them. In OFDM signals, it is conceivable that the over-sampling factor $M/K=8$ is sufficiently large to measure the PAPR. As the over-sampling factor, $M/K \geq 4$ is commonly used. The discussion about the over-sampling factor is described in \cite{sampling}.

In Section IV, we have discussed the PAPR of the filters obtained from the problem $(P)$. As seen in Fig. \ref{fig:papr}, the PAPR for the signals with filters obtained from the problem $(P)$ is higher than that with the Dolph-Chebyshev filter and the OFDM signals. Since each BER of the filters obtained by the problem $(P)$ is nearly the same, it is conceivable that each $\tilde{\sigma}$, which is defined in Section IV is the same. Therefore, the PAPR for the filters obtained by the problem $(P)$ is uniformly high and they are nearly the same. On the other hands, the PAPR for the Dolph-Chebyshev filter is lower than that for the filters obtained by the problem $(P)$. From this result and the result about BER, there is a relation between BER and PAPR in UF-OFDM systems, that is, PAPR gets higher as the SNR gets larger.

\section{Conclusion}
In this paper, we have shown the UF-OFDM model and the criteria of designing filters, which is related to side-lobes and SNR. Then, we have obtained the convex optimization problem and the filters as its solutions. As numerical results, we have shown their BER, power spectrum and PAPR and compared them with the Dolpf-Chebyshev filter and conventional OFDM signals. From these results, the BER for our filters is lower than that for the Dolpf-Chebyshev filter. In particular, the side-lobe of the filter whose parameter $\lambda=0.0001$ is the lowest. This filter is appropriate for communication systems since the power of side-lobes is low. However, the signals of UF-OFDM systems have high PAPR than ones of conventional OFDM systems. It is necessary to consider this relation for communication systems.


\section*{Acknowledgment}
The one of the authors, Hirofumi Tsuda, would like to thank Dr. Shin-itiro Goto for his advise.



%

\end{document}